# DIFFUSION THEORY OF ELECTRICAL CONTACT RESISTANCE OF THERMOELECTRIC SUPERLATTICE –METAL COUPLE

P.V.Gorskyi[1,2], DSc (phys-math), gena.grim@gmail.com

[1)]Institute of Thermoelectricity of the NAS and MES of Ukraine, [2)]Yu.Fedkovych Chernivtsi National University

*The paper proposes a diffusion theory of the electrical contact resistance of a thermoelectric superlattice (TES) - metal couple. On its basis, the thickness of the transient contact layer and the value of the electrical contact resistance of TES - metal are determined. Moreover, the law of growth of the transient contact layer is found. It is shown that the upper limit of the electrical contact resistance for bismuth telluride-nickel couple at 500 K can reach $4 \cdot 10^{-5} Ohm \cdot cm^2$. At the same time, it was found that the non-parabolicity of the band spectrum of the superlattice, described by the Fivaz model, depending on the degree of openness of the Fermi surface of the thermoelectric superlattice, can reduce this value by at least 60-100 times. However, the obtained experimental data, on the one hand, and the predicted significant difference in contact resistance for n- and p-type materials, on the other hand, give reason to assume that the electrical contact resistance of the thermoelectric material-metal for the bismuth-nickel telluride couple is due not so much to the diffusion of nickel in TES, as to the deviation surface of the semiconductor from the ideal plane and the presence of an oxide film on the surface of the TES.*

**Key words**: *diffusion, matter balance, boundary layer thickness, superlattice, contact resistance.*

*Introduction*

*The current state of the problem, the purpose and objectives of the research*. The metal-semiconductor contact resistance consists of two main parts: the resistance caused by the potential barrier at the metal-semiconductor interface and the resistance caused by the transient layer formed by the diffusion of the metal into the semiconductor and (or) vice versa. Regarding the "barrier" part of the resistance, it can be said that in the most general case it is caused by two mechanisms: thermoelectron emission and tunneling of charge carriers through the barrier. In [1], we gave a brief overview of these mechanisms and the conditions for their manifestation. Regarding the manifestation of these mechanisms in the real operating conditions of thermoelectric devices, it turned out that even in cooling modules on the hot side, and even more so in generator modules, there is a combined or transient mode in which both mechanisms operate. Based on such considerations, we calculated the "barrier" part of the contact resistance.

As for the part of the contact resistance due to the transient layer, it should be noted that prior to our studies carried out in the above mentioned work, this part was found by simply multiplying the thickness of the transient layer by the resistivity of the semiconductor material, in particular TEM. The thickness of the transient layer was determined experimentally based on the results of metallographic and electron microscopic studies. We, however, proposed to determine the distribution of the metal in the transient layer as a solution to the equation of stationary diffusion from a source with constant volume intensity. In this case, the thickness of the transient layer acted as a parameter of the theory, the limits of change of which were estimated according to the experimental data given in the literature. It should be noted that although we obtained a certain agreement between the results of our theoretical calculations and the experiment, this approach is suitable in the case when the contact structure is created, for example, by sputtering a contact metal or an anti-diffusion layer on the surface of a semiconductor from a source with constant volume intensity. In the case of creating contact structures by electrochemical or purely chemical deposition of an anti-diffusion layer on the TEM surface with subsequent soldering, the diffusion process, during which the transient layer is formed, is fundamentally nonstationary and the theory developed in [1, 2] ceases to be correct.

Experimental studies of the properties of contact structures and transient contact layers, analysis of their effect on consumer characteristics, failure and durability of thermoelectric devices, temporal evolution of transient contact layers, the effect of semiconductor surface treatment methods on electrical contact resistance, methods for reducing contact resistances and methods for their accurate measurement are discussed in a number of works, including [3-7]. But to date, a consistent theory of electrical resistance of contact structures created by soldering and their degradation has not been developed. A partial solution to this problem is the aim of this work. In the course of the work, the equation of non-stationary diffusion of metal in TEM is solved taking into account the matter balance, and the exact law of growth of the transient contact layer and the normalized distribution of the metal in it are found. The temperature dependence of the electrical contact resistance of the transient contact layer is calculated and the process of long-term diffusion of nickel in the TEM layer based on Bi(Sb)–Te(Se) is analyzed, which is of fundamental importance for the reliability of thermoelectric generator modules.

*Solution of the equation of non-stationary diffusion of metal in TEM and its consequences.* If the diffusion coefficient $D$ of metal in TEM is considered constant, then in a one-dimensional approximation this equation will be given by:

$$\frac{\partial c}{\partial t} = D \frac{\partial^2 c}{\partial x^2} \; , \qquad (1)$$

where $c$ is concentration of metal atoms, $t$ is time, $x$ is coordinate counted from the interface deep into TEM.

For the case of a semi-confined medium, the solution of equation (1) can be taken in the form:

$$c = c_0 \text{erfc}\left(x/\sqrt{2Dt}\right), \qquad (2)$$

where $\text{erfc}(...)$ is the so-called additional error integral. Solution (2) satisfies the obvious initial condition $c(t=0)=0$ and the obvious boundary condition $c(x=0)=c_0$. However, the solution of Eq. (1) in the form (2) does not yet determine the law of growth of the transient layer with time. In order to determine it, we equate the number of metal atoms that will leave the nickel boundary layer in a certain time $t_0$ due to the boundary flux to the number of them that will be distributed in the transient layer of thickness $x_0$ in the same time:

$$c_0 \frac{2\sqrt{Dt_0}}{\sqrt{\pi}} = c_0 \int_0^{x_0} \text{erfc}\left(x/2\sqrt{Dt_0}\right) dx. \qquad (3)$$

When writing the equation, we believe that only the thickness of the boundary layer of nickel changes during the diffusion process, and the concentration of atoms in it remains unchanged until it is completely exhausted. Eq.(3) leads to the following transcendental equation:

$$\frac{1}{\sqrt{\pi}} = \int_0^{\alpha_0} \text{erfc}(\alpha)d\alpha, \qquad (4)$$

the solution of which is $\alpha_0=6.771$, and, therefore, the growth of the transient layer occurs according to the law $x_0 = 13.552\sqrt{Dt_0}$. Thus, the thickness $h$ of the spent part of nickel layer is matched by the thickness $6.771\sqrt{\pi}h \approx 12h$ of transient contact layer. In this case, the concentration distribution of metal atoms in the transient layer, normalized to its thickness, is determined by the formula:

$$c(x) = c_0 \text{erfc}(6.771 x/x_0), \qquad (5)$$

and it is shown in Fig.1.

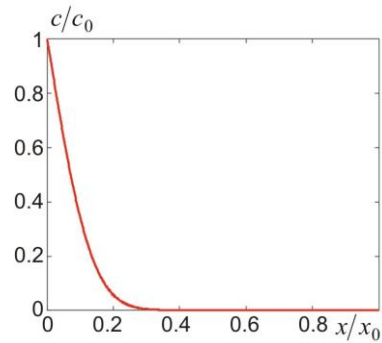

*Fig.1. Distribution of metal atoms in a transient contact layer*

It is noteworthy that the ratio between the thickness of the transient contact layer and the thickness of the spent nickel layer is an invariant of the diffusion process that determines the formation of the transient contact layer. The value of the diffusion coefficient determines only the time dependence of the thickness of the transient contact layer and, consequently, the contact resistance of the TEM-metal.

*Calculation of electrical contact resistance.* Using the theory of composites and taking into account the phenomenon of percolation associated with the formation of metal clusters in the thickness of the transient layer, we determine the electrical contact resistance TES - metal at an arbitrary temperature $T$. To do this, we first determine the volume fraction of metal in the transient layer:

$$\nu(x) = \frac{(A_m/\rho_m)\mathrm{erfc}(6.771x)}{(A_m/\rho_m)\mathrm{erfc}(6.771x) + (M_s/\rho_s)\mathrm{erf}(6.771x)}, \qquad (6)$$

where $A_m$, $\rho_m$ are atomic mass and metal density, respectively, $M_s$, $\rho_s$ are molecular mass and TES density, respectively, erf(…) and erfc(…) are the so-called error integral and additional error integral. Then, with regard to percolation effect, the electrical contact resistance of the transient layer will be determined as:

$$r_{ce}(T) = 48h \int_0^1 \{R(x) - \sigma_m(T)(3\nu(x)-1) - \sigma_s(T)(2 - 3\nu(x))\}^{-1} dx, \qquad (7)$$

where $h$ is the thickness of spent metal layer, $\sigma_m(T)$ is specific electrical conductivity of metal, $\sigma_s(T)$ is specific electrical conductivity of TES and, besides

$$R(x) = \sqrt{[\sigma_m(T)(3\nu(x)-1) + \sigma_s(T)(2 - 3\nu(x))]^2 + 8\sigma_m(T)\sigma_s(T)} \; . \qquad (7)$$

The temperature dependence of the specific electrical conductivity of a thermoelectric superlattice is determined as follows:

$$\sigma_s(T) = 0.08 \frac{\sqrt{\pi^5 n_0 a} e^2 l(T)}{ah T_{2D}(T)} \times \\ \times \int_0^\infty \int_0^\pi \frac{y \exp[(y + K^{-1}(1-\cos x) - \eta)/T_{2D}(T)] dx dy}{\{1 + \exp[(y + K^{-1}(1-\cos x) - \eta)/T_{2D}(T)]\}^2 \sqrt{2y + 4\pi K^{-1} n_0 a^3 \sin^2 x}}, \qquad (8)$$

where $T_{2D}(T) = 4\pi m^* kT/h^2 n_0 a$, $l(T) = l_0 T_0/T$, $K = h^2 n_0 a/4\pi m^* \Delta$, $n_0$ is volume concentration of free charge carriers, $m^*$ is effective mass of charge carriers in the plane of layers, $a$ is the smallest distance between translationally equivalent layers, $l_0$ is the length of free path of charge carriers at certain fixed temperature $T_0$, $\Delta$ is the half-width of a narrow miniband, which in the framework of the Fivaz model characterizes the interlayer motion of charge carriers, the rest of the designations are generally accepted. The scattering of charge carriers on the deformation potential of acoustic phonons was considered to be decisive. Normalized chemical potential $\eta$ maybe determined from equation

$$\frac{T_{2D}(T)}{\pi}\int_0^\pi \ln\left\{1+\exp\left[(\eta - K^{-1}(1-\cos x))/T_{2D}(T)\right]\right\}dx - 1 = 0. \quad (9)$$

Specific calculations were performed in the temperature range $T=300 – 500$ K for different values of the parameter $K$, which characterizes the degree of openness of the Fermi surface of the TES, and, therefore, the degree of non-parabolicity of its band spectrum.

The results of calculations of the temperature dependences of electrical contact resistance for the TES based on bismuth telluride — nickel couple at the thicknesses of spent nickel layer 5 and 20 µm, respectively, are given in Fig.2. For comparison, the same figure shows the results of calculations of the contact resistance of TES with a parabolic band spectrum. In the latter case, the temperature dependence of the effective mass caused by the electron-phonon interaction was taken into account. In all the cases the specific electrical conductivity of metal was considered to be inversely proportional to temperature and at $T_0=300$K it was assumed that $\sigma_{m0} = 1.43 \cdot 10^5$ Cm/cm.

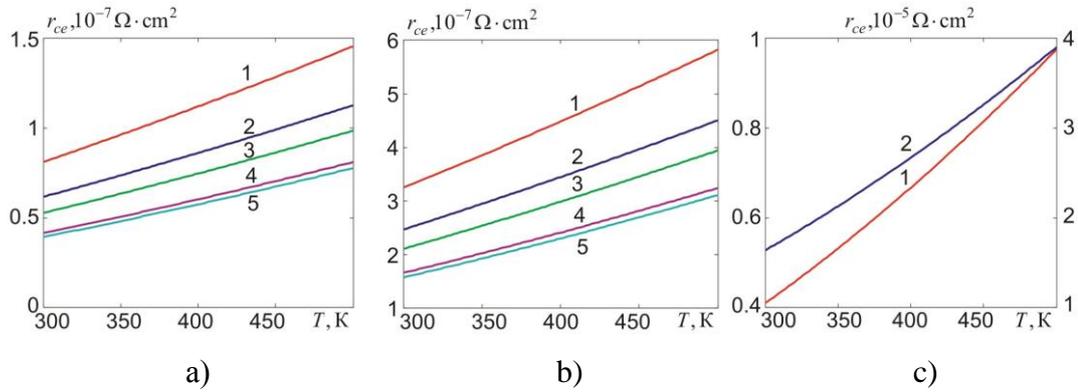

a)        b)        c)

Fig.2. Temperature dependences of the contact resistance of TES based on bismuth telluride-nickel couple at the thicknesses of spent nickel layer: a) 5 µm, b) 20 µm at the degrees of nonparabolicity $K$: 1 – 0.1, 2 – 0.5, 3 – 1, 4 – 5, 5 – 10, as well as for the parabolic case (c) at the thicknesses of spent nickel layer: 1 – 5 µm (left axis), 2 – 20 µm (right axis).

When constructing the plots in Fig. 2 a, b, the following values of TES parameters were taken: $m^* = m_0 = 9.1 \cdot 10^{-31}$ kg, $a = 3$ nm, $n_0 = 3 \cdot 10^{19}$ cm$^{-3}$, $l_0 = 20$ nm, $T_0=300$K.

When constructing the plots in Fig.2c it was assumed that the specific electrical conductivity of bismuth telluride at $T_0=300$K is $\sigma_{s0} = 1.4 \cdot 10^3$ Cm/cm and is inversely proportional to $T^{1.7}$. Moreover, when constructing all the plots it was assumed that $A_m = 58$, $\rho_m = 9100$ kg/m$^3$, $M_s = 802$, $\rho_s = 7860$ kg/m$^3$. It can be seen from the figures that at the thickness of spent nickel layer 5 μm the electrical contact resistance of TES in the considered range of temperatures and nonparabolicity degrees varies from $3.93 \cdot 10^{-8}$ to $1.46 \cdot 10^{-7}$ Ohm·cm$^2$, and at the thickness of spent nickel layer 20 μm — from $1.57 \cdot 10^{-7}$ to $5.82 \cdot 10^{-7}$ Ohm·cm$^2$. However, in the case of thermoelectric material with a parabolic band spectrum at the thickness of spent nickel layer 5 μm the contact resistance varies from $4.09 \cdot 10^{-6}$ to $9.75 \cdot 10^{-6}$ Ohm·cm$^2$, and at the thickness of spent nickel layer 20 μm – from $1.64 \cdot 10^{-5}$ to $3.9 \cdot 10^{-5}$ Ohm·cm$^2$. Traditionally, the anti-diffusion layer is made 20 μm thick. Therefore, a comparison of the calculated values of the electrical contact resistance with those traditionally "assigned" and used in the design of thermoelectric generators and coolers allows us to conclude that the thickness of the diffusing nickel layer does not exceed 5 μm, while the remaining 15 μm remain on the surface. In so doing, the transition from traditional contact structures to heterostructures described by the Fivaz model, depending on the degree of openness of the Fermi surface of TES, should reduce the contact resistance by a factor of 60-100.

It should be noted that, according to [8], the diffusion coefficients of nickel in *p* and *n* type TEMs differ even in orders of magnitude. This difference is caused primarily by the difference in activation energies of nickel diffusion in these materials. Therefore, the values of contact resistances reached during the same time of formation of the transient layer should differ significantly from each other. In practice, however, such a difference is not observed. Therefore, it should be assumed that the main role in the formation of contact resistance is played by the roughness of the contact surfaces of the thermoelectric legs and the presence of oxide films on them.

*Conclusions*

1. A strict diffusion theory of the electrical contact resistance of the transient layer is constructed.

2. The growth law of the transient contact layer is found.

3. The temperature dependence of the electrical contact resistance of the transient contact layer was calculated both for the case of TES and for the case of a thermoelectric material with a traditional band spectrum, and it was shown that the transition to TES makes it possible to reduce the electrical contact resistance by a factor of 60-100.